\documentclass[twocolumn,showpacs,preprintnumbers,amsmath,amssymb]{revtex4} 
\usepackage{epsfig}

\usepackage{graphicx}
\usepackage{dcolumn}
\usepackage{bm}

\usepackage{amsthm} 
 
\newcommand{\ket}[1]{\mbox{$ | #1 \rangle $}} 
\newcommand{\bra}[1]{\mbox{$ \langle #1 | $}}

\begin{document} 
 
 
\title{Upper bound on the secret key rate distillable from effective quantum correlations with imperfect detectors} 
 
\author{Tobias Moroder, Marcos Curty, and Norbert L\"{u}tkenhaus} 
\affiliation{Quantum Information Theory Group, Institut f\"{u}r Theoretische
  Physik I, and Max-Planck Research Group, Institute of Optics, Information
  and Photonics, Universit\"{a}t Erlangen-N\"{u}rnberg, Staudtstra{\ss}e 7/B2,
  91058 Erlangen, Germany} 
 
\date{\today} 
 
\begin{abstract} 
We provide a simple method to obtain an upper bound on the secret 
key rate that is particularly suited to analyze practical 
realizations of quantum key distribution protocols with imperfect 
devices. We consider the so-called \emph{trusted device scenario} 
where Eve cannot modify the actual detection devices employed by 
Alice and Bob. The upper bound obtained is based on the available 
measurements results, but it includes the effect of the noise and 
losses present in the detectors of the legitimate users. 
\end{abstract} 
 

\maketitle 
 
\section{INTRODUCTION} 
 
Quantum key distribution (QKD) \cite{Wiesner83,bennett84a} is a technique that
allows two parties, typically called Alice and Bob, to generate an
unconditionally secure secret key. This secret key can then be used together
with the Vernam cipher \cite{vernam} to achieve unconditionally secure
communications against any possible eavesdropper, named Eve. 
 
Typical practical QKD protocols distinguish two phases in order to generate a
secret key. In the first phase, an effective bipartite quantum state is
distributed between Alice and Bob. This state creates correlations between
them and it might contain as well hidden correlations with Eve. Next, Alice
and Bob perform a set of measurements on the given signal states in order to 
measure these correlations. As a result, they obtain a joint probability
distribution $p(a_i,b_j)\equiv p_{ij}$ describing their classical outcomes. In
the second phase, Alice and Bob try to distill a secret key from these
correlated data $p_{ij}$ by means of public discussions over an
authenticated classical channel. This procedure includes postselection of
data, error correction to reconcile the data, and privacy amplification to
decouple the data from Eve \cite{Norbert99}. In this second phase, no
quantum-mechanical manipulations are performed; it is a completely classical
process.  
 
Two types of schemes are used to create the correlated data in the first phase
of QKD. In {\it entanglement-based} (EB) schemes, a source, which is assumed
to be under Eve's control, produces a bipartite quantum state $\rho_{AB}$ that
is distributed to Alice and Bob. Eve could even have a third system entangled
with those given to the legitimate users. Alice and Bob measure each incoming
signal by means of two {\it positive operator valued measures} (POVM)
$\{A_i\}$ and $\{B_j\}$, respectively. While the subsystems measured by Alice
and Bob result in correlations described by a joint probability distribution
$p_{ij}$, Eve can use her subsystem to obtain information about the data of
the legitimate users.  
 
In {\it prepare and measure} (PM) schemes, Alice prepares a state
$\ket{\varphi_i}$ with probability $p_i$ and sends it to Bob. On the receiving
side, Bob measures each received signal with a POVM described by the quantum
operators $\{B_j\}$. Generalizing the ideas introduced by Bennett {\it et al.}
\cite{mermin}, the signal preparation process in PM schemes can be thought of
as follows: Alice produces first the bipartite quantum state
$\ket{\psi_{source}}_{AB}=\sum_i \sqrt{p_i} \ket{\alpha_i}_A\ket{\varphi_i}_B$
and, afterward, she measures the first subsystem in the orthogonal basis
$\ket{\alpha_i}_A$. This action generates the (nonorthogonal) signal states
$\ket{\varphi_i}$ with probabilities $p_i$. It is important to note that in PM
schemes, the reduced density matrix of Alice,
$\rho_A=\text{Tr}_B(\ket{\psi_{source}}_{AB}\bra{\psi_{source}})$, is fixed
and cannot be modified by Eve. The state $\rho_A$ depends only on the
probabilities $p_i$ and on the overlap of the signals states
$\ket{\varphi_i}$. This information is known to all the parties. To include it
in the measurement process one can add to the observables $\{A_i\otimes B_j\}$
measured by Alice and Bob other observables $\{C_k \otimes\openone\}$ such
that the observables $\{C_k\}$ form a tomographically complete set of Alice's
Hilbert space \cite{curty04suba}.  
 
From now on, we will always use the term entanglement and detection devices
without making any further distinction between these two different QKD
schemes. Moreover, $p_{ij}$ and $\{A_i\otimes B_j\}$ refer to the complete set
of measurements, i.e., they include also the observables $\{C_k
\otimes\openone\}$ for PM schemes.  
 
An essential question in QKD is whether the correlated data contained in
$p_{ij}$ allow Alice and Bob to generate a secret key. In
Ref. \cite{curty04a}, it has been proven that a necessary precondition for
secure key distribution is the provable presence of quantum correlations in
$p_{ij}$. In this context, see also Ref.~\cite{gisinwolf}. Given the set of
measurements performed by Alice and Bob together with the resulting
probabilities $p_{ij}$, the legitimate users can calculate the equivalence
class $\mathcal{S}$ of quantum states that are compatible with the observed
data $p_{ij}$:  
\begin{equation}\label{eq_class} 
\mathcal{S}=\left\{ \rho_{AB}\ |\ \text{Tr}(A_i \otimes B_j\ 
\rho_{AB})=p_{ij}, \ \forall i,j \right\}. 
\end{equation} 
By definition, every state $\rho_{AB} \in \mathcal{S}$ can represent the state
shared by Alice and Bob before their measurements \cite{temp1}. According to
Ref. \cite{curty04a}, in order to be able to distill a secret key from the
observed data, it is necessary to prove that the equivalence class
$\mathcal{S}$ does not contain any separable state. See also
Ref.~\cite{gisinwolf}. This task is called \emph{effective entanglement
  verification}, where the word ``effective'' is used to emphasize that the
entanglement which might be contained in the state $\rho_{AB}$ is destroyed by
Alice and Bob after their measurements or might not have been present at all
in PM schemes. To deliver this entanglement proof, one can employ any
separability criteria \cite{Separability}. One possibility particularly suited
for this purpose is the use of entanglement witnesses
\cite{curty04a,curty04suba}.  
 
From Ref. \cite{curty04a}, we learn that, once the first phase of QKD is
complete, we need to find quantum-mechanical correlations in $p_{ij}$. As it
is, the effective entanglement verification answers only partially the
important question of how much secret key can Alice and Bob obtain from their
correlated data: it just tells if the secret key rate $K_S$ is zero or it may
be positive. The problem of finding upper bounds for $K_S$ was recently
analyzed from an entanglement theory point of view by K. Horodecki {\it et
  al.} in Ref.~\cite{horodecki05a}. These authors showed that the value of
$K_S$ that can be extracted from a quantum state $\rho_{AB}$ can be strictly
greater than the distillable entanglement \cite{ben96}, and it is generally
bounded from above by the regularized relative entropy of entanglement
$E_r^{\infty}(\rho_{AB})$ \cite{horodecki05a,vedral}. This upper bound takes
only into account the form of $\rho_{AB}$ and assumes perfect detection
devices for Alice and Bob. To calculate the regularized relative entropy of
entanglement of a given quantum state is, in general, a quite difficult task,
and analytical expressions are only available for some particular states
\cite{aude}. Another well-known upper bound on $K_S$ is the so-called
intrinsic information proposed by Maurer and Wolf \cite{maurer99a}. See also
Refs.~\cite{curty04a,christandl04a,acin05a}. In this case, one can easily
incorporate the possible imperfections of the detectors in the upper
bound. However, it requires to solve an infimum ranging over the set of all
possible discrete probability distributions, or discrete channels, which is in
principle not easy to compute. (See also Ref.~\cite{chris}.)  
 
In this paper, we present a simple method to obtain an upper bound on $K_S$
that is particularly suited to evaluate the secret key rate on practical
realizations of QKD schemes. Moreover, this procedure has the advantage that
is is straightforward to calculate. It is based on the correlated data
$p_{ij}$, and it also monitors any form of detector
imperfections. Specifically, we consider the so-called \emph{trusted device
  scenario} where Eve cannot modify the actual detection devices employed by
Alice and Bob. We assume that the legitimate users have complete knowledge
about their detectors, which are fixed by the actual experiment. These
detectors might be noisy and might introduce losses and they are characterized
by the POVMs $\{A_i\}$ and $\{B_j\}$.  
 
The paper is organized as follows. In Sec.~II, we introduce two desired
properties for any upper bound on $K_S$ for practical QKD: effective
entanglement verification, and the ability to monitor any kind of
imperfections in Alice's and Bob's detection setups. This section contains as
well a brief summary of some known upper bounds on $K_S$. Section III includes
the main result of the paper: There we introduce a straightforward method to
obtain an upper bound on $K_S$ that satisfies the conditions presented in
Sec.~II. This result is then illustrated in Sec.~IV for two well-known QKD
protocols with imperfect detectors: the four-state \cite{bennett84a} and the
six-state \cite{bruss98a} QKD schemes. Finally, Sec.~V concludes the paper
with a summary.

\section{Desired properties and known upper bounds on $K_S$}

\subsection{Desired properties}\label{sec2a} 
 
The secret key rate $K_S$, i.e., the rate of secret key that can be obtained
per signal state sent by Alice to Bob, is one of the most important figures of
merit in order to compare the performance of different QKD schemes. Ideally,
one would like to calculate the maximum achievable value of $K_S$ from the
data available once the first phase of the QKD protocol is completed. Note
that the second phase is just a classical procedure that is completely
independent of the actual experimental setup used.  
 
In a similar spirit, any upper bound on the achievable secret key rate $K_S$
should as well be valid for any arbitrary public communication protocol
performed during the second phase of QKD. The upper bound only depends on the
observed data $p_{ij}$ together with the particular POVMs $\{ A_i \}$ and $\{
B_j \}$ used by Alice and Bob.  
 
This gives rise to one main requirement expected from any upper bound on
$K_S$: if the observed data $p_{ij}$ can originate from a separable state,
then any upper bound must vanish \cite{curty04a}. That is, the upper bound
might be considered as the generalization of the necessary effective
entanglement verification.  
 
From a practical point of view, it would be necessary that any upper bound on
$K_S$ includes the effect of imperfect devices. Especially, it should be able
to take into account the low detection efficiency and the noise in the form of
dark counts introduced by current detection devices. This fact is of special
importance in order to compare different realizations of QKD, and it can be
used to see the tradeoff between desired and realizable implementations of
QKD.

\subsection{Known upper bounds on $K_S$} 
 
In this section, we review very briefly some known upper bounds on $K_S$ that
apply to particular scenarios:  
 
\textit{Mutual information $I(A;B)$:} Suppose Alice and Bob are connected by a
public channel and have access to repeated independent realizations of two
random variables, denoted as $A$ and $B$, and which are characterized by a
joint probability distribution $p_{AB}$. An upper bound on the secret key rate
$K_S$ is given by the mutual information $I(A;B)$
\cite{maurer93a,maurer99a}. This quantity is defined in terms of the Shannon
entropy $H(X)=-\sum_{x\in X} p(x)\log{p(x)}$ and the Shannon joint entropy
$H(X,Y)=-\sum_{x\in X}\sum_{y\in Y} p(x,y)\log{p(x,y)}$ as  
\begin{equation} 
I(A;B) = H(A) + H(B) - H(A,B). 
\end{equation} 
This result is not surprising since the mutual information quantifies the
reduction in the uncertainty of the random variable $A$ because of the
knowledge of $B$.  
 
This upper bound can directly be used for the case of QKD, just by taking as
$p_{AB}$ the correlated data $p_{ij}$. Moreover, in this case, one can easily
incorporate the effect of the imperfections in the detectors when calculating
$I(A;B)$. Unfortunately, the mutual information can only provide an upper
bound on $K_S$ that is not really tight.  
 
\textit{Intrinsic information $I(A;B \downarrow E)$:} It provides an upper
bound on $K_S$ for a particular classical key-agreement scenario. First, we
describe the classical situation, and afterward, we adapt the upper bound to
the QKD scenario.  
 
In the classical case, Alice, Bob, and Eve have access to independent
realizations of three random variables, $A$, $B$, and $E$, and which are
described by the joint probability distribution $p_{ABE}$. The intrinsic
information, denoted as $I(A;B \downarrow E)$, constitutes an upper bound on
the secret key rate $K_S$ \cite{maurer99a}. The intrinsic information is
defined as 
\begin{equation}\label{int}  
I(A;B\downarrow{}E) = \inf_{E \to \bar{E}} I(A;B|\bar{E}), 
\end{equation} 
where the minimization runs over all possible classical channels $E \to
\bar{E}$ characterized by the transition probability $P_{\bar{E}|E}$, and
where $I(A;B|\bar{E})$ is the mutual information between Alice and Bob given
the public announcement of Eve's data based on the probabilities
$P_{AB\bar{E}}$. This quantity is defined in terms of the conditional Shannon
entropy $H(X|\bar{e})=\sum_{x\in X} - p(x|\bar{e}) \log_2 p(x|\bar{e})$ as  
\begin{equation} 
I(A;B|\bar{E}) = \sum_{\bar{e}\in \bar{E}} P(\bar{e}) 
\Big[H(A|\bar{e})+H(B|\bar{e})-H(A,B|\bar{e})\Big]. 
\end{equation} 
The intrinsic information satisfies \cite{maurer99a} 
\begin{equation} 
0\leq{}K_S\leq{}I(A;B \downarrow E)\leq{}I(A;B). 
\end{equation} 
That is, $I(A;B|\bar{E})$ is a tighter bound on $K_S$ than the 
mutual information. However, recently it has been proven that the 
secret key rate can be smaller than the intrinsic information 
\cite{renner03a,note2}. 
 
More important for QKD, the upper bound based on the intrinsic information can
be adapted to the case where Alice, Bob, and Eve start sharing a tripartite
quantum state instead of a joint probability distribution. For this purpose,
one can consider all possible tripartite states that Eve can establish using
her eavesdropping method, and all possible measurements she could perform on
her subsystem. This gives rise to a set of possible extensions ${\cal P}$ of
the probability distribution $P_{AB}$ to $P_{ABE}$. Now one can define the
intrinsic information as \cite{curty04a}  
\begin{equation}\label{normar} 
I(A;B\downarrow{}E) = \inf_{{\cal P}}\;I(A;B|E) \; . 
\end{equation} 
 
As in the case of the mutual information, also this bound allows us to include
the imperfections of the detection devices when calculating
$I(A;B\downarrow{}E)$. Moreover, it provides effective entanglement
verification, i.e., $I(A;B\downarrow{}E)=0$ if and only if the equivalence
class $\mathcal{S}$ contains a separable state
\cite{curty04a,acin05a}. Unfortunately, it requires us to solve an infimum
problem that is not easy to compute. Note that if the range of the random
variable $E$ is finite, then the infimum becomes a minimum over channels with
the same alphabet \cite{chris}.   
 
\textit{Regularized relative entropy of entanglement $E_r^\infty(\rho_{AB})$:}
Suppose Alice and Bob share several copies of a quantum state $\rho_{AB}$, and
they are allowed to perform arbitrary local operations and classical
communication (LOCC). The regularized relative entropy of entanglement
$E_r^\infty(\rho_{AB})$ is an upper bound on $K_S$ \cite{horodecki05a}. The
relative entropy of entanglement $E_r(\rho_{AB})$ is given by
\cite{vedral,vedral2}  
\begin{equation} 
E_r(\rho_{AB})=\inf_{\sigma_{sep}}\ Tr[\rho_{AB}\left( 
\log\rho_{AB}-\log\sigma_{sep}\right)]. 
\end{equation} 
where the infimum is taken over all separable states $\sigma_{sep}$. The
regularized version of $E_r(\rho_{AB})$ is given by \cite{donald,aude}:  
\begin{equation} 
E_r^\infty(\rho_{AB})=\lim_{n \to \infty} 
\frac{E_r(\rho_{AB}^{\otimes n})}{n}. 
\end{equation} 
This quantity depends only on the shared quantum states $\rho_{AB}$, and
therefore it does not include the possible imperfections on the detectors of
Alice and Bob. Moreover, to calculate the regularized relative entropy of
entanglement of a given quantum state is, in general, a quite difficult task,
and analytical expressions are only available for some particular states
\cite{aude}.  
 
{\it Intrinsic information of a tripartite quantum state $I(\rho_{ABE})$:}
Assume Alice, Bob, and Eve share several copies of a quantum state
$\rho_{ABE}$, and they are allowed to perform arbitrary LOCC and may
communicate via a public channel. The intrinsic information of the tripartite
quantum state $\rho_{ABE}$ is defined as \cite{christandl04a}  
\begin{equation}
\label{int_t} 
I(\rho_{ABE})=\inf_{E_k} \sum_k p(e_k) S(A;B)_{e_k}, 
\end{equation} 
with $S(A;B)_{e_k}=S(\rho_A^{e_k}) + S(\rho_B^{e_k}) - S(\rho_{AB}^{e_k})$
being the quantum mutual information of the conditional state
$\rho_{AB}^{e_k}=Tr_E(E_k \rho_{ABE})/p(e_k)$, and represents an upper bound
on $K_S$ \cite{christandl04a}. Here $S(\rho)$ denotes the von Neumann entropy
of the state $\rho$. It is defined as
$S(\rho)\equiv-\text{Tr}(\rho\log_2\rho)$. The infimum given in
Eq.~(\ref{int_t}) runs over all possible POVMs $\{E_k\}$. 
 
This upper bound depends only on the quantum state $\rho_{ABE}$ and its
definition does not include the effect of possible imperfections in the
detectors of Alice and Bob in the most general case.

\section{Upper bound on $K_S$} 
 
In this section, we introduce a simple procedure, cf. Sec.~\ref{res}, to obtain
an upper bound on $K_S$ that satisfies the two desired conditions presented in
Sec.~II. Moreover, this procedure has the advantage that it is straightforward
to calculate. In order to do that, we start by presenting a naive method which
allows the derivation of a simple upper bound on $K_S$. This method is based
on imposing a particular eavesdropping strategy by Eve, and it guarantees that
any resulting upper bound is able to monitor possible detector inefficiencies
by construction. The necessary entanglement verification condition is then
included as a particular example of this method. This is done by selecting a
special eavesdropping strategy that exploits the best separable approximation
(BSA) \cite{lewenstein97a,karnas01a}.

\subsection{Simple method to derive upper bounds on $K_S$}\label{method} 
 
The idea is simple: just impose some \emph{particular} eavesdropping
strategy for Eve, and then use one of the already known upper bounds. The
upper bound obtained represents an upper bound for \emph{any} possible
eavesdropping strategy. If we use as starting point, for instance, the
intrinsic information, then the bound would be able to include the effect of
imperfect detectors. The method can be described with the following three
steps.  
 
(1) Select a \emph{particular} eavesdropping strategy for Eve. This strategy
    is given by the choice of a tripartite quantum state $\rho_{ABE}$ and a
    POVM $\{E_k\}$. The only restriction here is $\text{Tr}_{E}(\rho_{ABE})\in
    \mathcal{S}$. That is, the chosen strategy cannot alter the observed
    distribution $p_{ij}$.  
 
(2) Select three random variables $A$, $B$, and $E$ with probability
    distribution $p_{ijk}=\text{Tr}(A_i B_j E_k\ \rho_{ABE})$. Note that the
    POVMs of Alice $\{ A_i \}$ and Bob $\{ B_j \}$ are known and cannot be
    modified by Eve.  
 
(3) Calculate the intrinsic information $I_{E_k}(A;B \downarrow E)$, where the
    subscript $E_k$ denotes the chosen measurement strategy for Eve. The
    secret key rate $K_S$ is upper bounded by $I_{E_k}(A;B \downarrow E)$.  
 
Step (3) could also be substituted by an optimization taken over all possible
measurement strategies performed by Eve. That is, one might use as upper bound
\cite{acin05a}  
\begin{equation} 
K_S\leq\inf_{E_k} I_{E_k}(A;B \downarrow E) = \inf_{E_k} 
I_{E_k}(A;B | E). 
\end{equation} 
Moreover, in this last case, it is easy to see that the same upper bound could
also be obtained from the intrinsic information of a tripartite quantum
state. The corresponding steps for this case are included in App. \ref{ap_A}.  
 
This method does not always guarantee effective entanglement verification, but
this depends on the eavesdropping strategy selected in step (1). In
Sec.~\ref{eve} we introduce a special eavesdropping strategy that guarantees
that this property is always satisfied. It is based on the use of the BSA of a
quantum state \cite{lewenstein97a,karnas01a}. Next, we introduce the
equivalence between the BSA idea and effective entanglement verification.

\subsection{Best separable approximation and necessary entanglement
  verification}   
 
The best separable approximation (BSA) \cite{lewenstein97a,karnas01a} of a
given state $\rho_{AB}$ is the decomposition of $\rho_{AB}$ into a separable
state $\sigma_{sep}$ and an entangled state $\rho_{ent}$, while maximizing the
weight of the separable part. That is, given an arbitrary state $\rho_{AB}$,
it can be proven that this state can always be written in a unique way as  
\begin{equation} 
\rho_{AB} = \lambda_{max}(\rho_{AB}) \sigma_{sep} + 
[1-\lambda_{max}(\rho_{AB})] \rho_{ent}, 
\end{equation} 
where the entanglement state $\rho_{ent}$ has no product vectors in its range,
and the real parameter $\lambda_{max}(\rho_{AB})\geq{}0$ is maximal
\cite{lewenstein97a,karnas01a}.  
 
Given an equivalence class $\mathcal{S}$ of quantum states, one defines the
maximum weight of separability within the equivalence class,
$\lambda_{max}^{\mathcal{S}}$, as  
\begin{equation} 
\label{lambda_max} \lambda_{max}^{\mathcal{S}} = \max \left \{ 
\lambda_{max}(\rho_{AB})\ |\ \rho_{AB} \in \mathcal{S} \right \}. 
\end{equation} 
This parameter is related to the task of effective entanglement 
verification by the following observation. 
 
\textit{Observation 1}. Assume that Alice and Bob can perform local
measurements with POVM elements $A_i$ and $B_j$, respectively, to obtain the
probability distribution of the outcomes $p_{ij}$ on the distributed quantum
state $\rho_{AB}$. Then the following two statements are equivalent: (1) The
correlations $p_{ij}$ can originate from a separable state. (2) The maximum
weight of separability $\lambda_{max}^{\mathcal{S}}$ within the equivalence
class of quantum states $\mathcal{S}$ compatible with the observed data
$p_{ij}$ satisfies $\lambda_{max}^{\mathcal{S}}=1$.  
 
\textit{Proof}. If $p_{ij}$ can originate from a separable state, then there
exists $\sigma_{sep}$ such as $\sigma_{sep}\in\mathcal{S}$. Moreover, we have
that any separable state satisfy $\lambda_{max}(\sigma_{sep})=1$. The other
direction is trivial. $\blacksquare$  
 
Let us define $\mathcal{S}_{max}$ as the equivalence class of quantum states
composed by those states $\rho_{AB}\in\mathcal{S}$ with maximum weight of
separability  
\begin{equation} 
\mathcal{S}_{max} = \left \{ \rho_{AB}\in\mathcal{S}\  |\ 
\lambda_{max}(\rho_{AB})=\lambda_{max}^{\mathcal{S}} \right \}. 
\end{equation}

\subsection{Eavesdropping strategy}\label{eve} 
 
Eve's eavesdropping strategy is completely characterized by selecting a
tripartite quantum state $\rho_{ABE}$ and a POVM $\{E_k\}$. We consider a pure
state $\rho_{ABE}=\ket{\Phi}_{ABE}\bra{\Phi}$ that is a purification of a
state $\rho_{AB}$ chosen from the equivalence class $\mathcal{S}_{max}$.  
 
We can write the separable part $\sigma_{sep}$ and the entangled part
$\rho_{ent}$ of the BSA of $\rho_{AB}$ as 
\begin{eqnarray} 
\sigma_{sep} &=& \sum_i q_i \ket{\phi_i}_A\bra{\phi_i} 
\otimes \ket{\varphi_i}_B\bra{\varphi_i}, \\ 
\rho_{ent} &=& \sum_i p_i \ket{\psi_i}_{AB}\bra{\psi_i}. 
\end{eqnarray} 
The tripartite state $\ket{\Phi}_{ABE}$ is then given by 
\begin{eqnarray}
\label{st} 
\ket{\Phi}_{ABE}=\sum_i \sqrt{\lambda_{max}^{\mathcal{S}} q_i} 
\ket{\phi_i}_A\ket{\varphi_i}_B\ket{e_i}_E + \nonumber \\ 
\sum_j \sqrt{(1-\lambda_{max}^{\mathcal{S}})p_j} 
\ket{\psi_j}_{AB}\ket{f_j}_E, 
\end{eqnarray} 
where the states $\{\ket{e_i}_E,\ket{f_j}_E\}$ form an orthogonal basis on
Eve's subsystem. It is important to note that in both kinds of QKD schemes, EB
schemes and PM schemes, Eve can have access to the state $\ket{\Phi}_{ABE}$
given by Eq.~(\ref{st}) \cite{curty04a}. In an EB scheme, this is clear since
Eve is the one who prepares the state $\rho_{AB}$ and who distributes it to
Alice and Bob. In the case of PM schemes, we need to show additionally that
the state $\ket{\Phi}_{ABE}$ can be obtained by Eve by interaction with Bob's
system only. In the Schmidt decomposition, the state prepared by Alice,
$|\psi_{source}\rangle_{AB}$, can be written as $|\psi_{source}\rangle_{AB} =
\sum_i c_i |u_i\rangle_A |v_i\rangle_B$. Then the Schmidt decomposition of
$\ket{\Phi}_{ABE}$, with respect to system $A$ and the composite system $BE$,
is of the form $\ket{\Phi}_{ABE}= \sum_i\ c_i |u_i\rangle_A
|\tilde{e}_i\rangle_{BE}$, since $c_i$ and $|u_i\rangle_A$ are fixed by the
known reduced density matrix $\rho_A$ to the corresponding values of
$|\psi_{source}\rangle_{AB}$. Then one can find a suitable unitary operator
$U_{BE}$ such that $|\tilde{e}_i\rangle_{BE}=U_{BE}|v_i\rangle_B|0\rangle_E$
where $|0\rangle_E$ is an initial state of an auxiliary system.  
 
As a measurement strategy for Eve, we consider that she is restricted to use a
special class of measurements. This class of measurements can be thought of as
a two step procedure.  
 
(1) First, Eve distinguishes contributions coming from the separable part and
    from the entangled part of $\rho_{AB}$: $\sigma_{sep}$ and $\rho_{ent}$,
    respectively. This corresponds to a projection of Eve's subsystem onto the
    orthogonal subspaces $\Pi_{sep}=\sum_i \ket{e_i}_E\bra{e_i}$ and
    $\Pi_{ent}=\sum_j \ket{f_j}_E\bra{f_j}$.  
 
(2) Afterward, she performs a refined measurement on each subspace
    separately. In the separable subspace, Eve can obtain complete
    information, and no secret key can be distilled by Alice and Bob
    \cite{curty04a,acin05a}. This corresponds to a projection onto the
    orthogonal quantum states $\{\ket{e_i}\}$. In the entanglement part, Eve
    performs a POVM denoted as $\{F_l\}$.  
 
We find that Eve's measurement result belongs to the separable subspace with
probability $\lambda_{max}^{\mathcal{S}}$, and therefore, $K_S=0$.  
 
With probability $1-\lambda_{max}^{\mathcal{S}}$, Eve's subsystem is in the
entanglement subspace and $K_S$ might be bigger than zero. After the first
step of Eve's measurement, the conditional quantum state of Alice, Bob, and
Eve, denoted as $\rho_{ABE}^{ent} = \ket{\Phi_{ent}}_{ABE} \bra{\Phi_{ent}}$,
corresponds to a purification of $\rho_{ent}$, i.e.,  
\begin{equation} 
\ket{\Phi_{ent}}_{ABE}=\sum_j \sqrt{p_j} 
\ket{\psi_j}_{AB}\ket{f_j}_E. 
\end{equation} 
 
Note that both steps can be described as well together by the following POVM:
$\{\tilde E_{sep}^i,\tilde E_{ent}^l\}$, with $\tilde
E_{sep}^i=\ket{e_i}\bra{e_i}$ and $\tilde E_{ent}^l=F_l \Pi_{ent}$.  
 
In the next section, we provide an upper bound for $K_S$ that arises from this
special eavesdropping strategy. Moreover, as we will see, the upper bound
obtained is straightforward to calculate.

\subsection{Resulting upper bound}\label{res} 
 
Only the entangled part $\rho_{ent}$ which appears in the BSA decomposition of
a given state $\rho_{AB}$ might allow Alice and Bob to distill a secret key in
the eavesdropping strategy proposed in the previous section. Moreover, Eve can
always find such an eavesdropping strategy for any $\rho_{AB} \in
\mathcal{S}_{max}$. This fact motivates the definition of a new equivalence
class of quantum states $\mathcal{S}_{max}^{ent}$  
\begin{equation} 
\mathcal{S}_{max}^{ent}= \left\{ \rho_{ent}(\rho_{AB})\ |\ 
\rho_{AB} \in \mathcal{S}_{max}) \right\}, 
\end{equation} 
where $\rho_{ent}(\rho_{AB})$ denotes the entangled part in the 
BSA of the state $\rho_{AB}$. 
 
\textit{Theorem 1}. Consider all possible bipartite entanglement states
$\rho_{ent}\in\mathcal{S}_{max}^{ent}$, and consider all possible POVMs
$\{F_l\}$ that Eve could perform on a purification $\ket{\Phi_{ent}}_{ABE}$ of
the quantum state $\rho_{ent}$. This gives rise to a set $\mathcal{P}$ of
tripartite probability distributions $p_{ijl}=\text{Tr}( A_i B_j F_l
\ket{\Phi_{ent}}_{ABE}\bra{\Phi_{ent}})$, where $\{A_i\}$ and $\{B_j\}$
represent the POVMs measured by Alice and Bob. The secret key rate $K_S$ is
upper bounded by  
\begin{equation} 
  \label{bound} 
  K_S \leq (1-\lambda_{max}^{\mathcal{S}}) \inf_{\mathcal{P}} I^{ent}(A;B|E), 
\end{equation} 
where $I^{ent}(A;B|E)$ represents the classical conditional mutual information
of three random variables distributed accordingly to $p_{ijl}$.  
 
\textit{Proof}: This proof is straightforward since, by construction,
Eq.~(\ref{bound}) is an upper bound of the intrinsic information defined in
Eq.~(\ref{normar}). Note that to obtain Eq.~(\ref{bound}), we assume a
particular type of eavesdropping for Eve, and in Eq.~(\ref{normar}), the
infimum is taken over all possible eavesdropping strategies. $\blacksquare$  
 
Instead of optimizing over all possible states
$\rho_{AB}\in\mathcal{S}_{max}^{ent}$ and all possible measurements performed
by Eve, one could also take any state in $\mathcal{S}_{max}^{ent}$ and
calculate the infimum of $I^{ent}(A;B|E)$ over all possible POVMs $\{F_l\}$
employed by Eve. This fact simplifies the calculation of the upper bound on
$K_S$.  
 
\textit{Corollary 1}. Given a state $\rho_{ent}\in\mathcal{S}_{max}^{ent}$,
the secret key rate $K_S$ is upper bounded by  
\begin{equation}\label{bound2} 
K_S \leq (1-\lambda_{max}^{\mathcal{S}}) \inf_{F_l} 
I_{F_l}^{ent}(A;B|E), 
\end{equation} 
with $I_{F_l}^{ent}(A;B|E)$ being the classical conditional mutual information
calculated on the probability distribution $p_{ijl}=\text{Tr}( A_i B_j F_l
\ket{\Phi_{ent}}_{ABE}\bra{\Phi_{ent}})$, and where $\ket{\Phi_{ent}}_{ABE}$
denotes a purification of $\rho_{ent}$.  
 
\textit{Proof}. Equation (\ref{bound2}) is an upper bound of
Eq.~(\ref{bound}). Note that in Eq.~(\ref{bound2}), we take a particular state
$\rho_{ent}\in\mathcal{S}_{max}^{ent}$, while the infimum in Eq.~(\ref{bound})
includes all possible states
$\rho_{ent}\in\mathcal{S}_{max}^{ent}$. $\blacksquare$  
 
The upper bounds provided by Theorem $1$ and Corollary $1$ are easier to
calculate than the one based on the intrinsic information defined in
Eq.~(\ref{normar}) \cite{simple}. However, they still demand solving a
difficult optimization problem. Next, we provide a simple upper bound on $K_S$
that is straightforward to calculate. Then, in Sec.~\ref{section_example}, we
illustrate the performance of this upper bound for two well-known QKD
protocols with imperfect detectors: the four-state \cite{bennett84a} and the
six-state \cite{bruss98a} QKD schemes. We compare it with the upper bound
given by the regularized relative entropy of entanglement \cite{horodecki05a}.
 
\textit{Corollary 2}: The secret key rate $K_S$ is upper bounded by 
\begin{equation}\label{bound3} 
K_S \leq (1-\lambda_{max}^{\mathcal{S}}) I^{ent}(A;B), 
\end{equation} 
where $I^{ent}(A;B)$ denotes the mutual information calculated on 
the probability distribution $\tilde p_{ij}=\text{Tr}( A_i B_j \rho_{ent})$
with $\rho_{ent}\in\mathcal{S}_{max}^{ent}$.  
 
\textit{Proof}: Equation (\ref{bound3}) is an upper bound of
Eq.~(\ref{bound2}), {\it i.e.}, $\inf_{F_l}
I_{F_l}^{ent}(A;B|E)\leq{}I^{ent}(A;B)$. Note that Eve could always select a
POVM with only one element $\tilde F_l=\openone$. In this case, we have
$I_{\tilde F_l}^{ent}(A;B|E)=I^{ent}(A;B)$. $\blacksquare$  
 
In Sec.~\ref{section_example}, we show that, for qubit-based QKD protocols, the
upper bound given by Corollary $1$ and by Corollary $2$ coincide.  
 
The main difficulty when evaluating this last upper bound for a particular
realization of QKD relies on obtaining $\lambda_{max}^{\mathcal{S}}$ and
$\rho_{ent}$. This can be solved by applying results from relaxation theory of
nonconvex problems \cite{Shor,Lasserre,Par}. See also Ref.~\cite{jens}. The
solution is included in App. \ref{ap_B}.  
 
The three bounds on $K_S$ obtained in this section satisfy the two desired
properties included in Sec.~\ref{sec2a}: the factor
$1-\lambda_{max}^{\mathcal{S}}$ accounts for effective entanglement
verification by Observation $1$, and the probabilities $p_{ijl}$ and $\tilde
p_{ij}$ reflect possible detector inefficiencies.

\section{Evaluation of the upper bound} 
\label{section_example}

In this section, we evaluate the upper bound on $K_S$ given by
Eq.~(\ref{bound3}) for two well-known qubit-based QKD protocols with imperfect
detectors: the four-state \cite{bennett84a} and the six-state \cite{bruss98a}
QKD schemes. We refer here to single-photon implementations of the qubit. The
state of the qubit is described by some degree of freedom in the polarization
of the photon.  
 
We start by describing the detection devices employed by Alice and Bob. They
are characterized by some noise in the form of dark counts which are, to a
good approximation, independent of the incoming signals, and by their
detection efficiency $\eta$.  
 
\textit{Dark counts:} For simplicity, we assume that only Bob's detection
device is affected by the presence of dark counts. We consider that Alice's
detectors are ideal. The total dark count probability is denoted as $d$. A
typical value for this parameter is $d=10^{-6}$. This scenario can be modeled
by transforming every ideal POVM element $B_j$ of Bob into a noisy element
$\tilde B_j$ given by  
\begin{equation} 
\tilde B_j = (1-d) B_j + d_j \openone_B, 
\end{equation} 
where the parameters $d_j$ satisfy $d=\sum_j d_j$. 
 
The effect of the noise can be included in the calculation of the upper bound
given by Eq.~(\ref{bound3}) via the classical mutual information
$I^{ent}(A;B)$. We have that $I^{ent}(A;B)>I^{ent}(A;\tilde B)$ simply because
the raw data is affected by more noise which needs to be corrected.  
 
It is important to note that the effective entanglement verification alone
cannot include the noise coming from the dark counts of the detectors. In the
ideal case, Alice and Bob observe outcomes governed by the ideal probability
distribution $p_{ij}$, which defines the equivalence class of quantum states
$\mathcal{S}$. In the noisy scenario, on the contrary, Alice and Bob obtain
the probability distribution $\tilde p_{ij}= (1-d) p_{ij} + d_j p_{i}$
defining an equivalence class $\tilde{\mathcal{S}}$. In principle, both
probability distributions are different. However, in the trusted device
scenario, Alice and Bob know all the parameters $d_j$. This means that they
can obtain $p_{ij}$ from $\tilde p_{ij}$ : $p_{ij}=1/(1-d)(\tilde p_{ij}-d_j
p_{i})$, and where $p_i=\sum_j \tilde p_{ij}$ \cite{ex}. That is, we obtain
$\tilde{\mathcal{S}}=\mathcal{S}$: any upper bound which only depends on the
shared quantum state $\rho_{AB}$ provides the same upper bound independently
of the noise introduced by the detectors.  
 
\textit{Detector efficiency:} Detectors are characterized as well by their
detector efficiency $\eta_j$. This effect can be modeled by a combination of a
beam splitter of transmittance $\eta_j$ and an ideal detector \cite{yurke}. A
typical value for $\eta_j$ in current realizations of QKD is approximately
$0.15$. In order to include the losses of the detectors, we can transform
every POVM element $B_j$ that corresponds to a ``click'' event into 
\begin{equation} 
  \tilde B_j = \eta_j B_j. 
\end{equation} 
Additionally, the event ``no click'' corresponds to the following 
operator: 
\begin{equation} 
  B_{vac} = \sum_j (1-\eta_j) B_j+\ket{\text{vac}}\bra{\text{vac}}. 
\end{equation} 
where $\ket{\text{vac}}$ represents the vacuum state. 
 
As in the case of dark counts, the effect of the losses is incorporated in
Eq.~(\ref{bound3}) via the classical mutual information $I^{ent}(A;B)$. For
simplicity, we consider that $\eta_j=\eta$ for all the detectors.  
 
\begin{figure} 
\centering 
\includegraphics[scale=0.57]{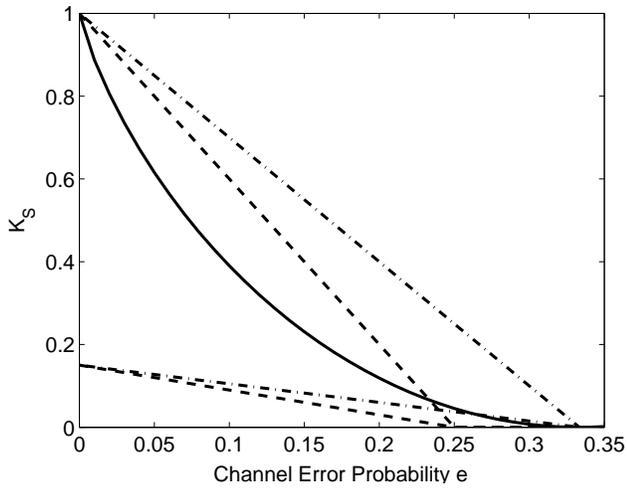} 
\caption{Upper bound on the secret key rate $K_S$ given by Eq.~(\ref{bound3})
  for the four-state (dashed) and the six-state (dash-dotted) QKD schemes. The
  classical correlated data $p_{ij}$ is obtained by measuring the quantum
  state $\rho_{AB}(e)=(1-2e) \ket{\psi^+}\bra{\psi^+}+e\openone{}/2$, where
  $e$ is the error probability of a depolarizing channel. The two upper lines
  (dashed and dash-dotted) correspond to the ideal case where the total dark
  count probability is $d=0$, and the detector efficiency is $\eta=1$. The two
  lines below represent the typical case in current realizations of QKD:
  $d=10^{-6}$ and $\eta=0.15$. The solid line of the graphic represent the
  upper bound for $K_S$ given by the regularized relative entropy of 
  entanglement $E_r^\infty(\rho_{AB}(e))$. Note that we assume an asymmetric 
  basis choice to suppress the sifting effect \cite{lo05a}.}  
  \label{figure_example} 
\end{figure}

In Fig. \ref{figure_example}, 
we illustrated the upper bound given by Eq.~(\ref{bound3}) for the four-state
\cite{bennett84a} and the six-state \cite{bruss98a} QKD schemes. In the case
of the four-state EB protocol, Alice and Bob perform projection measurements
onto two mutually unbiased bases, say the ones given by the eigenvectors of
the two Pauli operators $\sigma_x$ and $\sigma_z$. In the corresponding PM
scheme, Alice can use as well the same set of measurements but now on a
maximally entangled state. For the case of the six-state EB protocol, Alice
and Bob perform projection measurements onto the eigenvectors of the three
Pauli operators $\sigma_x, \sigma_y,$ and $\sigma_z$ on the bipartite qubit
states distributed by Eve. In the corresponding PM scheme Alice prepares the
eigenvectors of those operators by performing the same measurements on a
maximally entangled two-qubit state. Note that here we are not using the
general approach introduced previously, $\ket{\psi_{source}}_{AB}=\sum_i
\sqrt{p_i} \ket{\alpha_i}_A\ket{\varphi_i}_{B}$, to model PM schemes, since
for this protocol it is sufficient to consider that the effectively
distributed quantum state consists only of two qubits.  

We model the transmission channel as a depolarizing channel with error
probability $e$. This means that, in both protocols, the joint probability
distribution $p_{ij}$ is obtained by measuring the quantum state  
\begin{equation} 
\rho_{AB}(e)=(1-2e) \ket{\psi^+}\bra{\psi^+}+\frac{e}{2} \openone, 
\end{equation} 
where the state $\ket{\psi^+}$ represents a maximally entangled two-qubit
state: $\ket{\psi^+}=1/\sqrt{2}(\ket{00}+\ket{11})$. In
Fig. \ref{figure_example}, we assume as well that the total dark count
probability is $d=10^{-6}$, and the detector efficiency is
$\eta=0.15$. Moreover, we consider that all detectors have the same dark count
probability.  
 
It is important to note that in this simple two-qubit scenario, the
conditional mutual information $I_{F_l}^{ent}(A;B|E)$ reduces to
$I^{ent}(A;B)$: in this case $\rho_{ent}$ is just a pure entangled state
\cite{lewenstein97a}, {\it i.e.}, the purification $\ket{\Phi_{ent}}$ of the
quantum state $\rho_{ent}$ is of the form
$\ket{\Phi_{ent}}=\ket{\psi}_{AB}\otimes\ket{\varphi}_E$. For qubit-based QKD
protocols, therefore, the upper bound given by Corollary $1$ and by Corollary
$2$ coincide since $\inf_{F_l} I_{F_l}^{ent}(A;B|E)=I^{ent}(A;B)$. In order to
calculate the maximum weight of separability $\lambda_{max}^{\mathcal{S}}$,
and its associated entangled state $\rho_{ent}$ which are necessary to
evaluate Eq.~(\ref{bound3}) we use the method described in Appendix
\ref{ap_B}.  
 
Figure \ref{figure_example} includes also a comparison with the upper bound on
$K_S$ given by the regularized relative entropy of entanglement
\cite{horodecki05a}. To calculate $E_r^\infty(\rho_{AB}(e))$, we use the
results included in Ref. \cite{aude}. Despite its simplicity, the upper bound
given by Eq.~(\ref{bound3}) can provide a tighter bound on $K_S$ than the one
proposed in Ref.~\cite{horodecki05a}, when dealing with typical parameter
values for imperfect detectors. It must be mentioned here, however, that the
bound given by $E_r^\infty(\rho_{AB}(e))$ was proposed in a different
scenario, and the possibility of having noisy and lossy detectors was not
considered. The results obtained can also be compared with the best lower
bounds for the tolerable error rate $e$ arising from known security proofs
\cite{Chau}: $e=0.2$ for the case of the four-state protocol, and $e=0.276$
for the six-state protocol.

\section{Conclusion} 
 
A necessary precondition for secure quantum key distribution (QKD) is that
sender and receiver can use their available measurement results to prove the
presence of entanglement in a quantum state that is effectively distributed
between them. Moreover, this result applies both to prepare and measure and
entanglement-based schemes.  
 
Unfortunately, this effective entanglement verification answers only partially
the important question of how much secret key can be obtained by Alice and Bob
from their correlated data: it just tells if the secret key rate is zero or it
may be positive.  
 
In this paper, we present a simple method to obtain an upper bound on the
secret key rate that is particularly suited to evaluate practical realizations
of QKD schemes. It is based on the correlated data, but it also monitors any
form of detector imperfections. In particular, we consider the so-called
\emph{trusted device scenario}, where Eve cannot modify the actual detection
devices employed by Alice and Bob. We assume that the legitimate users have
complete knowledge about their detectors, which are fixed by the actual
experiment.

\section{Acknowledgments} 
 
The authors wish to thank M. Horodecki, M. Christandl, W. Mauerer, J. Rigas,
G. O. Myhr, J. M. Renes, and K. Tamaki for very useful discussions. This 
work was supported by the DFG under the Emmy Noether programme, 
the European Commission (Integrated Project SECOQC).

\appendix 
 
\section{Equivalent method}\label{ap_A} 
 
In this Appendix, we present an alternative method to the one introduced in
Sec.~\ref{method} that provides exactly the same upper bound on $K_S$. It uses
the intrinsic information of a tripartite quantum state, and it can be
described with the following steps:  
 
(1) Select a \emph{particular} eavesdropping strategy for Eve. This strategy
    is given by the choice of a tripartite quantum state $\rho_{ABE}$. The
    restriction here is $\text{Tr}_{E}(\rho_{ABE})\in \mathcal{S}$.  
 
(2) Compute the so-called \textit{ccq} state. This state is of the form
    \cite{devetak}: 
    \begin{equation} 
    \rho_{ABE}^{ccq}=\sum_{ij} \ket{ij}_{AB}\bra{ij} \otimes \rho_E^{ij} 
    \end{equation} 
    where the state $\rho_E^{ij}$ is not normalized and it is given by 
    $\rho_E^{ij}=\text{Tr}_{AB}(A_i B_j \rho_{ABE})$. 
 
(3) The upper bound on $K_S$ is given by the intrinsic information of the
    tripartite \textit{ccq} state $\rho_{ABE}^{ccq}$.  
 
The equivalence of this method and the one introduced in Sec.~\ref{method} can
be seen as follows: if the chosen tripartite state $\rho_{ABE}$ is the same
for both methods and one fixes a measurement strategy $\{ E_k \}$ for Eve,
then the conditional quantum mutual information $S(A;B)_{e_k}$ is equal to
$I_{E_k}(A;B|E=e_k)$. Note that in this case, the conditional state becomes
$\rho_{AB}^{ccq,e_k}=\sum_{ij} p(a_i,b_j|e_k) \ket{ij}_{AB}\bra{ij}$.

\section{Obtaining $\lambda_{max}^{\mathcal{S}}$ and $\rho_{ent}$}\label{ap_B} 
 
Here we use the method proposed in Ref.~\cite{jens} to obtain the parameter
$\lambda_{max}^\mathcal{S}$ and the density operator $\rho_{ent}$. It is based
on results from relaxation theory of nonconvex problems
\cite{Shor,Lasserre,Par}, notably the method of Lasserre \cite{Lasserre}. The
central observation in Ref.~\cite{jens} is that many problems related with
entanglement can be cast into the form of optimization problems with
polynomial constraints of low degree (three or less). The polynomial part of
the optimization problems is still nonconvex and computationally expensive to
solve. However, one can find hierarchies of solutions in a way that each step
is a better approximation to the exact solution than the previous
one. Moreover, each step can be efficiently solved via semidefinite
programming \cite{Semi}. The hierarchy is asymptotically complete, in the
sense that the exact solution is asympotically attained.  
 
It is important to note that the method introduced in Ref.~\cite{jens} is not
only meant as a numerical method, but each instance of the hierarchy delivers
a semidefinite program that is accessible with analytical methods.  
 
Next, we explain how to cast the problem of finding
$\lambda_{max}^\mathcal{S}$ and $\rho_{ent}$ into the desired form analyzed in
Ref.~\cite{jens}.  
 
The equivalence class of quantum states $\mathcal{S}$ is defined by the POVMs
$\{ A_i \otimes B_j \}$ and the observed data $p_{ij}$: $\rho_{AB} \geq 0$
belongs to $\mathcal{S}$ if it satisfies $\text{Tr}(A_i \otimes B_j \rho_{AB})
= p_{ij}$ for all $i,j$. The BSA for $\rho_{AB}$ can be written in the
following way  
\begin{equation} 
\rho_{AB} = \min_{\text{Tr}(\tilde \rho_{ent})} \tilde \sigma_{sep} + 
\tilde \rho_{ent} 
\end{equation} 
with $\tilde \sigma_{sep} \equiv \lambda_{max}(\rho_{AB}) \sigma_{sep}$ and
$\tilde \rho_{ent} \equiv [1-\lambda_{max}(\rho_{AB})] \rho_{ent}$. Separable
states $\tilde \sigma_{sep}$ can be characterized in terms of product vectors
$\tilde \sigma_{sep}=\sum_i P_i$, with $P_i = p_i \ket{\psi_i}\bra{\psi_i}
\otimes \ket{\phi_i}\bra{\phi_i}$. To guarantee that the operators $P_i$ have
the desired product form, they must satisfy the following constraints
\cite{jens}:  
 
\begin{eqnarray} 
\text{Tr}[\text{Tr}_I(P_i)^2] &=& [\text{Tr}(P_i)]^2\\ 
P_i &\geq& 0, 
\end{eqnarray} 
with $I=\{A,B\}$. The associated optimisation problem can now be 
written as 
\begin{eqnarray}\label{op} 
\text{minimize}&& t, \nonumber\\ 
\text{subject to} && t \geq 1-\text{Tr}(\tilde \sigma_{sep}), \nonumber\\ 
  && \rho_{AB} \geq 0, \nonumber\\ 
  && \text{Tr}(\rho_{AB}) = 1, \nonumber\\ 
  && \text{Tr}(A_i \otimes B_j \rho_{AB}) = p_{ij}\ \forall i,j, \nonumber\\ 
  && \tilde \sigma_{sep} = \sum_i P_i, \nonumber\\ 
  && \text{Tr}[\text{Tr}_I(P_i)^2] = [\text{Tr}(P_i)]^2 \ \forall I=A,B,
  \nonumber\\  
  && P_i \geq 0, \nonumber\\ 
  && \rho_{AB} - \tilde \sigma_{sep} \geq 0, 
\end{eqnarray} 
where the parameter $t$ represents the trace of the quantum state $\tilde
\rho_{ent}$. This polynomial optimization problem can be solved with the help 
of Lasserre's method \cite{Lasserre}. For these calculations, the package
\texttt{GloptiPoly} \cite{gpmanual} based on \texttt{SeDuMi} \cite{SeDuMi} is
freely available. The package \texttt{GloptiPoly} has a number of
desirable features, in particular, it provides a certificate for
global optimality. 
 
Note that in low dimensional cases ($2\otimes2$, $2\otimes3$) the
characterization of separable states can be simplified to those states $\tilde
\sigma_\text{sep} \geq 0$ such as $\tilde \sigma_\text{sep}^{T_B} \geq 0$
\cite{peres}, where $T_P$ is the partial transposition, that is, the
transposition with respect to one subsystem. The problem given by
Eq.~(\ref{op}) can be reduced then to one containing only linear and
semidefinite constraints. Problems of this form can be solved very efficiently
with standard semidefinite programming modules \cite{Semi}.

\bibliographystyle{apsrev} 
\bibliographystyle{apsrev} 

\begin{thebibliography}{99}
\bibitem{temp1} The equivalence class $\mathcal{S}$ reduces to the trivial
  one, {i.e.}, it contains only one element up to a global phase, when the
  measurements realized by Alice and Bob provide complete tomographic
  information about $\rho_{AB}$. This is the case, for instance, of the
  six-state QKD protocol \cite{bruss98a}. Otherwise, $\mathcal{S}$ contains
  always more than one quantum state.
\bibitem{simple} Note that the rank of the quantum states involved in the
  calculation decrease, i.e.,
  $\text{rank}[\rho_{ent}(\rho_{AB})]\leq{}\text{rank}(\rho_{AB})$. Moreover,
  we have that $\text{rank}[\rho_{ent}(\rho_{AB})] \leq (N-1)(M-1)$, where $N$
  and $M$ represent the dimensions of Alice's and Bob's Hilbert space,
  respectively \cite{lewenstein97a}.
\bibitem{ex} In PM schemes, all the probabilities $p_i$ are known since the
  beginning. They can be obtained directly from the reduced density matrix of
  Alice $\rho_A$ and the POVM elements $A_i$, which are fixed.
\end{thebibliography}
 
\end{document}